\documentclass[aip,jcp,reprint,amsmath,amssymb,floatfix,citeautoscript]{revtex4-1}
\usepackage{amsmath}
\usepackage[usenames,dvipsnames]{color}
\usepackage{amssymb}
\usepackage{braket}
\usepackage{graphicx}
\usepackage{ragged2e}
\usepackage[labelsep=period,format=plain,justification=justified,font=footnotesize]{caption}
\DeclareCaptionJustification{myjust}{\justifying}
\usepackage{times}
\usepackage{txfonts}

\begin{document}

\title{Path integral approach to the Wigner representation of canonical density operators for discrete systems coupled to harmonic baths}
\author{Andr\'{e}s Montoya-Castillo\footnote{Corresponding Author}}
\email{am3720@columbia.edu}
\affiliation{Department of Chemistry, Columbia University, New York, New York, 10027, USA}
\author{David R. Reichman}
\email{drr2103@columbia.edu}
\affiliation{Department of Chemistry, Columbia University, New York, New York, 10027, USA}

\date{\today}

\begin{abstract}
 We derive a semi-analytical form for the Wigner transform for the canonical density operator of a discrete system coupled to a harmonic bath based on the path integral expansion of the Boltzmann factor. The introduction of this simple and controllable approach allows for the exact rendering of the canonical distribution and permits systematic convergence of static properties with respect to the number of path integral steps. In additions, the expressions derived here provide an exact and facile interface with quasi- and semi-classical dynamical methods, which enables the direct calculation of equilibrium time correlation functions within a wide array of approaches. We demonstrate that the present method represents a practical path for the calculation of thermodynamic data for the spin-boson and related systems. We illustrate the power of the present approach by detailing the improvement of the quality of Ehrenfest theory for the correlation function $\mathcal{C}_{zz}(t) = \mathrm{Re}\langle \sigma_z(0)\sigma_z(t)\rangle$ for the spin-boson model with systematic convergence to the exact sampling function.   Importantly, the numerically exact nature of the scheme presented here and its compatibility with semiclassical methods allows for the systematic testing of commonly used approximations for the Wigner-transformed canonical density.   
\end{abstract}

\maketitle

\twocolumngrid


\normalsize

\section{Introduction}

Practical and accurate representations of fully correlated canonical density operators are essential for the determination of both thermodynamic and dynamic properties of many-body systems. The description of the thermodynamics of a system provides access to quantities like entropy, heat capacity, and various susceptibilities, which provide insight into, for example, the nature of equilibrium phase transitions.  On the dynamical side, equilibrium time correlation functions, which require sampling from the full equilibrium Boltzmann operator, lie at the heart of the description of linear and nonlinear spectroscopy,\cite{PrinciplesNonlinearSpec} the determination of transport coefficients in condensed phase systems \cite{KuboStatMechII, VanKampen}, and the calculation of chemical rate constants \cite{Yamamoto1960,Chandler1978,Voth1989a, Voth1989}. The development of schemes that accurately represent the canonical density operator has been the objective of a large number of theoretical efforts that have, in turn, produced an impressive spectrum of numerically exact \cite{BinderMonteCarlo, Barker1979, Parrinello1984, Wilson1975,Bulla2008, White2009,Barthel2009,Schollwock2011, Tanimura2014,Tanimura2015} and approximate methods \cite{Cao1994a, Cao1994b, Jang1999,Craig2004, Craig2005b, Miller2001a, Miller2009}. However, despite significant progress, the calculation of static and dynamical properties of many-body quantum systems remains a challenging task.
 
For many complex systems, the phase space formulation of quantum mechanics, as encoded by the Wigner distribution, has provided a particularly convenient platform for the investigation of both dynamics and thermodynamics \cite{Wigner1932,Imre1967,Heller1976,Hillery1984a,Polkovnikov2010}.  While the phase space formulation provides a rigorous, if generally impractical, protocol for the evolution of operators via the Moyal bracket \cite{Imre1967,Hillery1984a}, its utility lies in its compatibility with the semi-classical hierarchy of techniques. The incorporation of the Wigner approach into these approximate methods not only sidesteps the complications associated with the Moyal bracket expansion, but also allows for the choice of the level of sophistication and accuracy necessary for dynamical calculations.   Indeed, this is an essential factor as the simple Ehrenfest \cite{Gerber1982, Stock1995}, surface hopping \cite{Tully1971,Tully1990}, and linearized semiclassical initial value representation \cite{Wang1998b, Sun1998, Shi2003d} (LSC-IVR) schemes become the only practical approaches for many complex systems.   In addition, the phase space framework has also been an integral component in the development of successful hybrid schemes that combine numerically exact quantum approaches or traditional perturbation theories with classical time evolution \cite{Wang2001a, Thoss2001, Berkelbach2012, Berkelbach2012b, Montoya2015a}.

 Unfortunately, the Wigner transformation of the canonical density for complex systems can rarely be obtained analytically, and its numerical determination contends with the challenge of the highly oscillatory phase associated with the Fourier transform \cite{Wang1998b}.  Nevertheless, a variety of approximations have been developed. These range from the simple replacement of the quantum Boltzmann operator with its classical counterpart, an approximation that is only appropriate at sufficiently high temperatures where the zero-point energy is negligible, to sophisticated path integral-based techniques \cite{Egorov1998, Skinner2001a, Kim2002a, Poulsen2003,Smith2015, Shi2003c, Frantsuzov2003, Liu2006,Liu2007, Liu2009, Liu2011a, Liu2014, Marinica2006, Beutier2014, Basire2013a, Bose2015}. These approaches have proven useful in the investigation of, for instance, vibrational spectra and relaxation rates \cite{Shi2003b, Skinner2001a, Lawrence2005, Poulsen2005}, proton transfer problems \cite{Basire2013a}, and quantum diffusion in para-hydrogen \cite{Poulsen2004} and liquid neon \cite{Liu2006}. The benefits of these approximations notwithstanding, the general accuracy of approximate Wigner transformed density operators in complex systems has been difficult to assess, especially when used in conjunction with dynamical calculations.  
 
Here we show that for impurity-type problems where the system-bath coupling is linear in the bath coordinates and the bath can be approximated as harmonic, the Wigner transform of the canonical density operator can be obtained analytically. For this reason, we focus on the simplest nontrivial model that captures the relaxation and dephasing of generic quantum systems coupled to a quantum bath with arbitrary coupling strength: the spin-boson (SB) model.  Of course, even for the SB model the integration of the bath degrees of freedom required by the Wigner transformation cannot be achieved without a Hamiltonian splitting procedure, achieved in the path integral framework by means of the Trotter approximation.  Naturally, the formalism provided here is not restricted to the SB model, but is also applicable to any generalization where the bath remains harmonic and the coupling linear in the bath coordinate. To ensure that the resulting density operator can be used in conjunction with quasi-classical methods such as the Ehrenfest and surface hopping schemes as well as with conventional semi-classical methods, we implement only the partial Wigner transform with respect to the bath degrees of freedom.  Extension to the full Wigner transformation can be achieved simply through use of the mapping variable \cite{Miller1978, Meyer1979, Stock1997, Thoss1999} or coherent state \cite{Gazeau2009} formalisms. 
 
 Importantly, the present scheme provides a computationally simple approach to the calculation of thermodynamic data for SB-type systems.  By providing a numerically exact representation for the initial conditions used in dynamical simulations, this method also represents a important benchmark for the use of approximate Wigner transformed canonical densities both in the static and dynamic contexts. This property allows us to demonstrate that the current method converges rapidly with respect to the number of path integral slices for a large region of parameter space, and that proper rendering of the canonical density can dramatically influence the accuracy of both thermodynamic and dynamic quantities.  Interestingly, the analytical expression derived here reveals the canonical density as a  linear superposition of bath distributions with weights determined by the paths allowed in configuration space.  
 
 It bears remarking that Moix, Zhao, and Cao have previously developed a related and highly efficient approach based on the influence functional formalism for the calculation of the reduced density matrix of a system coupled linearly to a harmonic bath \cite{Moix2012}.  The reduced density matrix, which corresponds to the partial trace over the bath degrees of freedom of the full canonical density operator, permits the calculation of thermodynamic averages of any \textit{system} operator, but precludes calculation of any non-system property.  In contrast, by providing an analytical form for the \textit{full} canonical density operator, our approach permits the calculation of \textit{any} thermodynamic average, albeit at a higher computational cost.  Another advantage of the present work is that, as stated above, it can be used to efficiently and exactly sample the initial conditions required for quasi- or semi-classical calculations of equilibrium time correlation functions. Similar to the work of Moix \textit{et al.}, our work can be easily generalized to $N$-level systems coupled to harmonic baths and is not limited to any specific form of the spectral density, $J(\omega)$.
 
The paper is organized as follows. In Sec.~\ref{Ch6Sec:Theory}, we introduce the formalism used in the paper.  Specifically, in Sec.~\ref{Ch6Subsec:PhaseSpace}, we present a brief review of the phase space formulation of quantum mechanics.  Sec.~\ref{Ch6Subsec:Hamil} introduces the SB Hamiltonian.  In Sec.~\ref{Ch6Subsec:CanonicalDensity} we outline the derivation of the Wigner-transformed canonical density for the SB model (the extended derivation can be found in Appendix \ref{Ch6App:PITreatmentRho}). Sec.~\ref{Ch6Sec:Results} contains the results and in Sec.~\ref{Ch6Sec:Conclusions} we conclude.  
 
\section{Theory}
\label{Ch6Sec:Theory}

\subsection{Phase Space Formulation}
\label{Ch6Subsec:PhaseSpace}

As stated in the Introduction, the phase space formulation of quantum mechanics provides a framework that integrates the use of Monte Carlo sampling of initial conditions coupled with trajectory-based methods associated with quasi- and semi-classical approaches.  Within this framework, the trace over two operators can be expressed in phase space as
	\begin{equation}\label{Ch6Eq:TracePhaseSpaceDef}
	\mathrm{Tr}[\hat{A}\hat{B}] = [2\pi ]^{-f} \int d\mathbf{x}d\mathbf{p}\ A^W(\mathbf{x}, \mathbf{p})B^W(\mathbf{x}, \mathbf{p}), 
	\end{equation}
where $A^W(\mathbf{x}, \mathbf{p})$ and $B^W(\mathbf{x}, \mathbf{p})$ are Wigner transformed versions of operators $\hat{A}$ and $\hat{B}$, which become functions of the classical coordinate and conjugate momentum variables $\mathbf{x}$ and $\mathbf{p}$, respectively, and $f$ is the number of degrees of freedom with respect to which the Wigner transform is performed.  The Wigner transform of an operator, $\hat{O}$, is defined as,
	\begin{equation}\label{Ch6Eq:WignerTransform}
	O^W(\mathbf{x}, \mathbf{p})  = \int d\mathbf{s}\ e^{-i \mathbf{p} \cdot \mathbf{s}} \bra{\mathbf{x} + \mathbf{s}/2} \hat{O} \ket{\mathbf{x} - \mathbf{s}/2}.
	\end{equation}

As Eqs.~(\ref{Ch6Eq:TracePhaseSpaceDef}) and (\ref{Ch6Eq:WignerTransform}) suggest, the phase space formulation can be used to obtain static averages when both $A$ and $B$ in Eq.~(\ref{Ch6Eq:TracePhaseSpaceDef}) are independent of time, or correlation funtions when at least one of the operators is time evolved.  In the following, we will be particularly interested in equilibrium time correlation functions of the form, 
	\begin{equation}\label{Ch6Eq:CFDefinition}
	\begin{split}
	\mathcal{C}_{AB}(t) &= \mathrm{Tr}[\rho A(0) B(t)]\\
	&= [2\pi \hbar]^{-f}\int d\mathbf{x}d\mathbf{p}\ [\rho A(0)]^W(\mathbf{x}, \mathbf{p}) [B(t)]^W(\mathbf{x}, \mathbf{p}),
	\end{split} 
	\end{equation}	 
	where $\rho = e^{-\beta H} / \mathrm{Tr}[e^{-\beta}H]$ is the canonical density operator, $\beta = [k_BT]^{-1}$ is the inverse of the thermal energy, and $B(t) = e^{iHt/\hbar}Be^{-iHt/\hbar}$. 
	
	The Wigner transform for products of operators (e.g., $[\rho A(0)]^W$ in Eq.~(\ref{Ch6Eq:CFDefinition})) may be expressed as,
	\begin{equation}
	\begin{split}
	[\hat{O}\hat{P}]^W(\mathbf{x}, \mathbf{p}) &= O^W(\mathbf{x}, \mathbf{p}) e^{\hbar \overset\leftrightarrow{\Lambda}/ 2i} P^W(\mathbf{x}, \mathbf{p}),
	\end{split}
	\end{equation}
where $\overset\leftrightarrow{\Lambda}$ is the Poisson bracket operator
	\begin{equation}
	\overset\leftrightarrow{\Lambda} = \overleftarrow{\nabla}_{\mathbf{p}} \cdot \overrightarrow{\nabla}_{\mathbf{x}} - \overleftarrow{\nabla}_{\mathbf{x}} \cdot \overrightarrow{\nabla}_{\mathbf{p}}
	\end{equation}
and the arrows above the gradient operators indicate the direction in which they act. For notational simplicity, we henceforth set $\hbar = 1$.  As a final note, we remark on the fact that it is well-known that the Wigner transform of the density operator need not be positive definite \cite{Hillery1984a, Polkovnikov2010}. This potential complication presents no difficulties in the calculations that follow.

\subsection{Hamiltonian}
\label{Ch6Subsec:Hamil}

The formalism we develop here is applicable to Hamiltonians consisting of a finite number of discrete states coupled to a noninteracting harmonic bath, with the coupling assumed to be linear in the bath coordinate. The reason for these restrictions is that the current treatment relies on the influence functional approach, which formally eliminates the bath degrees of freedom in the path integral framework \cite{Feynman1963, FeynmanHibbs}.

While this restriction may seem severe, it is noteworthy that a wide spectrum of problems in the condensed phase may be mapped to such a Hamiltonian.  For instance, the discrete degrees of freedom often correspond to a limited subset of the electronic or excitonic manifold coupled to an environment, often idealized as an infinite set of harmonic oscillators.  Such Hamiltonians can be written as a sum of system, bath, and coupling contributions, $H = H_S + H_B + H_{SB}$.  Perhaps the simplest in this class of models is the SB Hamiltonian \cite{Leggett1987, Weiss}.  In the SB model, the system part consists of two discrete states, 
	\begin{equation}
	H_S = \varepsilon\sigma_z + \Delta\sigma_x, 
	\end{equation}	 
where $\sigma_i$ corresponds to the $i^{th}$ Pauli matrix, $2\varepsilon$ is the bias energy difference between the two states, and $\Delta$ represents the off-diagonal coupling between the two sites and is assumed to be static. 

The bath consists of independent harmonic oscillators, 
 	\begin{equation}\label{Ch6Eq:HB}
	H_B = \frac{1}{2} \sum_k \Big[\hat{P}_k^2 + \omega_k^2 \hat{Q}_k^2 + \frac{c_k^2}{\omega_k^2}\Big],
	\end{equation}
	where $P_k$, $Q_k$ and $\omega_k$ are the mass-weighted momenta, coordinates, and frequency for the $k^{th}$ harmonic oscillator, respectively. The last term on the right hand side of Eq.~(\ref{Ch6Eq:HB}) is a constant term added for later convenience. As mentioned above, the system-bath coupling term is assumed to be linear in the bath coordinates and antisymmetric with respect to the system, 
	\begin{equation}\label{Ch6Eq:HSB}
	H_{SB} = \alpha \sigma_z \sum_k c_k \hat{Q}_k,
	\end{equation}
	where $c_k$ is the coupling constant describing the strength of the interaction between the system and the $k^{th}$ oscillator, and $\alpha = \pm 1$.  The spectral density, $J(\omega)$, fully determines the coupling between the system and the bath and is  assumed to take the functional form, 
	\begin{equation}
	\begin{split}
	J(\omega) &= \frac{2}{\pi}\sum_k \frac{c_k^2}{\omega_k}\delta(\omega - \omega_k),\\
	&= \frac{\pi}{2}\xi \omega e^{-\omega / \omega_c} \label{Ch6Eq:OhmicSD},
	\end{split}
	\end{equation}
where the cutoff frequency $\omega_c$ determines the correlation time for the bath at finite temperatures, and the Kondo parameter, $\xi$, is a dimensionless measure of the coupling between the system and bath.  The Kondo parameter is also proportional to the reorganization energy of electron transfer theory, $\lambda = \xi \omega_c / \pi = \pi^{-1} \int_0^{\infty} d\omega\ J(\omega)/\omega$, which  represents the energy dissipated after the system undergoes a Frank-Condon transition.  The functional form for the spectral density in the second line of Eq.~(\ref{Ch6Eq:OhmicSD}) corresponds to the often used Ohmic spectral density \cite{Leggett1987} with an exponential cutoff. We remark, however, that the approach presented here is not limited to any particular form of the spectral density.
 
\subsection{Canonical density: A path integral treatment} 
\label{Ch6Subsec:CanonicalDensity}

Referring back to Eq.~(\ref{Ch6Eq:CFDefinition}), it is clear that an expression for $\rho^W$ is necessary.  Because the system part of the Hamiltonian consists of discrete states, $\{ \ket{0}, \ket{1} \}$, we focus on deriving an expression for an arbitrary matrix element of the canonical density after a partial Wigner transform with respect to the bath degrees of freedom, 
 	\begin{equation}\label{Ch6Eq:SchematicRho}
	\begin{split}
	\rho^W_{a,b} &= [2\pi]^{-f}\int d\mathbf{s}\ e^{-i\mathbf{p}\cdot \mathbf{s}/\hbar} \rho_{a,b}(\mathbf{x} + \mathbf{s}/2, x - \mathbf{x}/2)\\
	&\equiv N_{ab}\cdot \mathcal{R}_{a,b}^W(\mathbf{x}, \mathbf{p}), 
	\end{split}
	\end{equation}
where $\rho_{a,b}(\mathbf{x} + \mathbf{s}/2, x - \mathbf{x}/2) = \bra{\mathbf{x} + \mathbf{s}/2} \bra{a}\rho\ket{b} \ket{\mathbf{x} - \mathbf{s}/2}$,  $a, b \in \{0,1\}$, $N_{ab}$ is a temperature dependent normalization constant, and $\mathcal{R}_{a,b}^W(\mathbf{x}, \mathbf{p})$ is a bath operator of unit trace, i.e., $\int d\mathbf{x} d\mathbf{p}\ \mathcal{R}_{a,b}^W(\mathbf{x}, \mathbf{p}) = 1$, which can be interpreted as the bath distribution function.  We henceforth drop the dependence of the bath distribution function on the bath coordinates and momenta, $(\mathbf{x}, \mathbf{p})$, for notational clarity.  We also note that we have included the prefactor $[2\pi]^{-f}$ in the definition of the Wigner transform of the canonical density so that it obeys the normalization condition $\sum_a \int d\mathbf{x}d\mathbf{p}\ \rho_{a,a}^W(\mathbf{x}, \mathbf{p}) = 1$.

For systems where the total Hamiltonian can be partitioned into two components that are simple to diagonalize, the path integral framework can provide a convenient route for obtaining the exponentiated form for the Hamiltonian necessary for the calculation of propagators and the Boltzmann factor.  In this case, we employ the separation adopted previously by Makri and coworkers in the development of the quasi-adiabatic path integral scheme \cite{Makri1992, Makarov1994, Makri1995a, Makri1995b}, $H = H_{ad} + H_{na}$, where $H_{ad} = H_{S}$ and $H_{na} = H_{B} + H_{SB}$, which refer to the adiabatic and nonadiabatic components of the Hamiltonian.   With this partitioning, we rewrite the Boltzmann factor using the Trotter factorization as an $N$-membered product of basic path integral units
	\begin{equation}\label{Ch6Eq:Trotter}
	e^{-\beta H} = \lim_{N\rightarrow \infty} [e^{-\beta H_{na}/2N}e^{-\beta H_{ad}/N}e^{-\beta H_{na}/2N}]^N.
	\end{equation}
	When $N$ is finite, the above equality ceases to be exact and the error it incurs is of the order $\mathcal{O}(N \cdot \exp\{ -\beta [H_{ad}, H_{na}] / 2N \})$. Also note that the Hermiticity of the Boltzmann factor is maintained by the symmetrical splitting in Eq.~(\ref{Ch6Eq:Trotter}).  
	Using the Trotter decomposition in Eq.~(\ref{Ch6Eq:Trotter}), introducing resolutions of the identity in the system and bath subspaces, $\mathbf{1}_S = \sum_{a} \ket{a}\bra{a}$ and $\mathbf{1}_B = \int d\mathbf{q} \ket{\mathbf{q}}\bra{\mathbf{q}}$, and performing the integrations over the bath coordinates analytically, it is possible to obtain expressions for temperature-dependent (global) normalization factor and bath distribution in Eq.~(\ref{Ch6Eq:SchematicRho}). After integration over the bath degrees of freedom, the sequence of spin-variables that characterize the path integral trajectory in configuration space remain, i.e., the sets $\{k_0, k_1, ..., k_N\}$ where $k_j \in \{0,1\}$. To illustrate this, consider the simpler case of treating the isolated subsystem Boltzmann factor via the path integral procedure (with $N =3$) such that 
	\begin{equation}
	\begin{split}
	\bra{k_3}e^{-\beta H_S}\ket{k_0} &\approx \sum_{k_1, k_2} \bra{k_3}e^{-\beta H_S/3}\ket{k_2} \times ...\ \times\\
	&\qquad \ \ \bra{k_2}e^{-\beta H_S/3}\ket{k_1} \bra{k_1}e^{-\beta H_S/3}\ket{k_0},
	\end{split}
	\end{equation}
where $\ket{k_j} \in \{\ket{0}, \ket{1} \}$. In the following, we refer to individual realizations of the sequence $\{k_0, k_1, k_2, k_3 \}$ as ``paths''.  Using this notation, 
	\begin{equation}\label{Ch6Eq:BathDistribution}
	\begin{split}
	\mathcal{R}_{a,b}^W(\mathbf{x}, \mathbf{p}) &= \mathcal{N} \sum_{\{ k_1, ..., k_{N-1} \}} \frac{\tilde{W}_{a,b}}{\mathcal{W}_{a,b}} \\
	& \prod_{l = 1}^{F} \exp\Big[-\gamma_p^{(l)}( p_l + i\tilde{\kappa}_p^{(l)})^2 -\gamma_x^{(l)} (x_l + \tilde{\kappa}_x^{(l)})^2 \Big],
	\end{split}
	\end{equation}
	\begin{equation}\label{Ch6Eq:GlobNormFactor}
	\begin{split}
	N_{ab} &=\frac{\mathcal{W}_{a,b}}{\sum_a  \mathcal{W}_{a,a}},
	\end{split}
	\end{equation}
	where $\mathcal{N}$ is a normalization factor, the ratio $\tilde{W}_{a,b}/\mathcal{W}_{a,b}$ corresponds to the weighting factors associated with individual paths, $\tilde{\gamma}_p$ and $\tilde{\gamma}_p$ ($i\tilde{\kappa}_p^{(l)}$ and $\tilde{\kappa}_x^{(l)}$) are the path-dependent variances (means) for the Gaussian distributions of coordinate and momentum of the $l^{th}$ oscillator, $p_l$ and $x_l$, respectively.  In this notation, the tilde denotes that a quantity is path-dependent. Detailed expressions for these quantities and their derivation can be found in Appendix \ref{Ch6App:PITreatmentRho}.
		
	The interpretation of Eq.~(\ref{Ch6Eq:BathDistribution}) is straightforward. The bath distribution function for the SB- and other impurity-type models where the bath is harmonic and the system-bath coupling linear in the bath coordinate can be expressed as a linear combination of Gaussian distributions in the bath coordinates where each contribution is weighted by a temperature- and path-dependent quantity $\tilde{W}_{a,b}/\mathcal{W}_{a,b}$ and for which the average displacements of the bath coordinate and momentum are also path-dependent quantities.  Importantly, Eqs.~(\ref{Ch6Eq:BathDistribution}) and (\ref{Ch6Eq:GlobNormFactor}) constitute the main result of the analytical manipulations presented in this work. We emphasize as well that the expressions for the canonical density of the SB model derived here may be used to calculate thermodynamic properties and averages and can be easily incorporated into a quasi- and semi-classical descriptions of the equilibrium time correlation functions.    This result is also to be considered in light of related treatments of the density operator, in particular the thermal Gaussian approximation \cite{Liu2006a} and the Feynman-Kleinert linearized path integral (FK-LPI) treatment\cite{Poulsen2003}.  In both, the Wigner transformed density operator is expressed as a single function rather than a superposition of Gaussian distributions.

\section{Results}
\label{Ch6Sec:Results}

	In this section we present some representative results obtained using Eqs.~(\ref{Ch6Eq:BathDistribution}) and (\ref{Ch6Eq:GlobNormFactor}) for thermodynamic averages of spin variables and dynamic calculations of the correlation function, $\mathcal{C}_{zz}(t) = \mathrm{Re}\langle\sigma_z(0)\sigma_z(t) \rangle$. The dynamics are calculated using the quasi-classical Ehrenfest method, which propagates the system (bath) variables in the time-dependent mean-field of the bath (system) and can be associated with an expansion of the Moyal operator $e^{\hbar\overset\leftrightarrow{\Lambda}/ 2i}$ to first order in $\hbar$ \cite{Grunwald2009}. Via comparison with numerically exact results for $\mathcal{C}_{zz}(t)$, we illustrate the sensitivity of the Ehrenfest dynamics to the accurate rendering of the canonical distribution.  Appendix \ref{Ch6App:Ehrenfest} provides details regarding the implementation of the Ehrenfest method. 
	
\begin{figure}
\centering
\includegraphics[width=8.5cm]{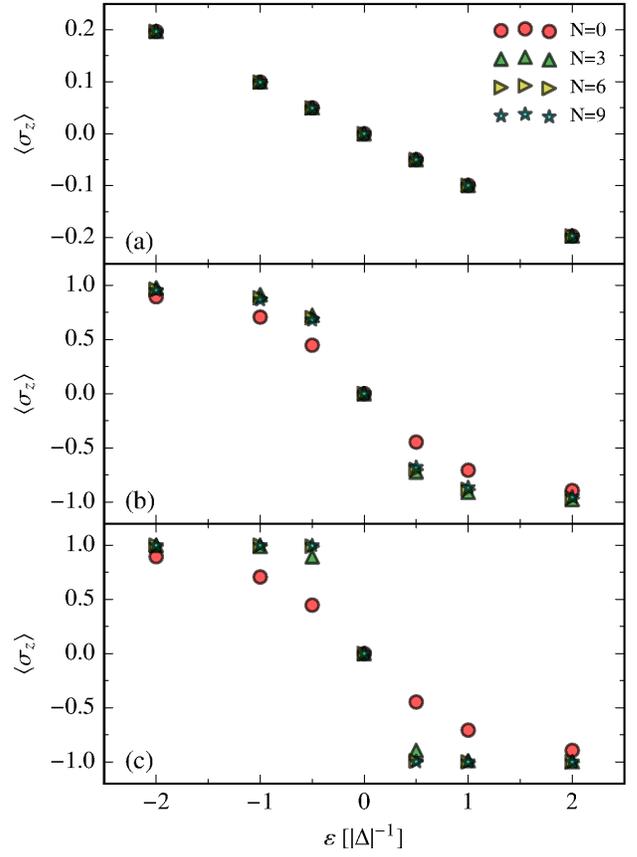} 
\caption{Calculation of the equilibrium population difference, $\langle \sigma_z \rangle$, as a function of the applied bias for the SB mode where $\Delta = \omega_c = 1$ and $\alpha = -1$.  For panel (a), $\beta = 0.1$, $\xi = 0.1$; for panel (b), $\beta = 5.0$, $\xi = 1.0$; for panel (c), $\beta = 10.0$, $\xi = 5.0$.  The different markers correspond to the use of different number of path integral slices in the thermodynamic calculation. }\label{Ch6Fig:WignerFig1}
\end{figure}
	
Before turning to dynamical calculations, we show some representative calculations of thermodynamic averages of the population difference at equilibrium, $\langle \sigma_z \rangle$, for different realizations of the SB model.  Fig.~\ref{Ch6Fig:WignerFig1} illustrates the convergence of $\langle \sigma_z \rangle$ with the number of path integral steps for three cases where $\beta$ or $\xi$ is increased.  As is evident from panel (a), $N = 0$ path integral slices is sufficient to obtain converged results in the high temperature, weakly coupled case.  As panels (b) and (c) indicate, with decreasing temperature and increasing system-bath coupling, the number of path integral slices necessary for the converged calculation of thermodynamic averages increases. This is consistent with the fact that the error associated with the Trotter decomposition is of order $\mathcal{O}(N \cdot \exp\{ -\beta [H_{ad}, H_{na}] / 2N \})$, where the contribution from $[H_{ad}, H_{na}]$ generally grows with increasing $\xi$. Remarkably, even for significantly lower temperature and stronger system-bath coupling ($\beta = 10.0$ and $\xi = 5.0$), $N = 6$ is sufficient to obtain converged results.  It is also worth noting that with increasing $\beta$ and $\xi$, the polarization of the SB model with net bias becomes more severe, as is indicated by the difference in the magnitude of polarization from panel (a) to (b) and (c), and with the faster onset of full polarization with $|\varepsilon|$ between panels (b) and (c). 

\begin{figure}
\centering
\includegraphics[width=8.5cm]{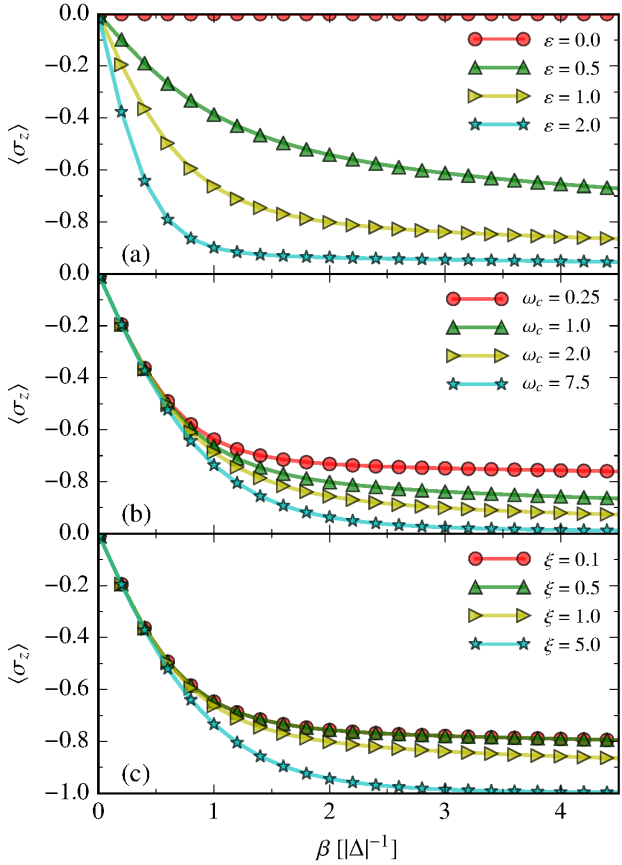} 
\caption{Expectation value for the equilibrium population difference as a function of inverse temperature $\beta$ and variation in the applied bias $\varepsilon$, characteristic response time of the bath $\omega_c$, and system-bath coupling strength $\xi$.  For all panels, $\Delta = 1$ and $\alpha = -1$.  For panel (a), $\omega_c = \xi = 1.0$; for $\varepsilon = \xi = 1.0$; and for (c), $\varepsilon = \omega_c  = 1.0$.}\label{Ch6Fig:WignerFig2}
\end{figure}

The current path integral approach to the density operator also permits the facile investigation of the dependence of thermodynamic averages on the continuous variation of parameters.  Fig.~\ref{Ch6Fig:WignerFig2} shows the dependence of the population difference as a function of $\beta$ with the variation of the applied bias $\varepsilon$, the characteristic frequency of the bath $\omega_c$, and the coupling between the system and bath $\xi$.  Consistent with physical intuition, panel (a) shows that the system becomes more polarized at equilibrium with increasing bias and favors the polarized state with decreasing temperature. The dependence of the results on the variation of the characteristic frequency of the bath shown in panel (b) indicates that the faster the response of the bath (larger $\omega_c$), the easier it becomes for the system to reach a stable polarized state, corresponding to the formation of a polaron.  Finally, panel (c) shows the dependence of the polarization on the system-bath coupling.  The results in panels (b) and (c) also agree with physical intuition which indicates that fast baths and strong system-bath coupling promote polaron formation.  

\begin{figure}
\centering
\includegraphics[width=8.5cm]{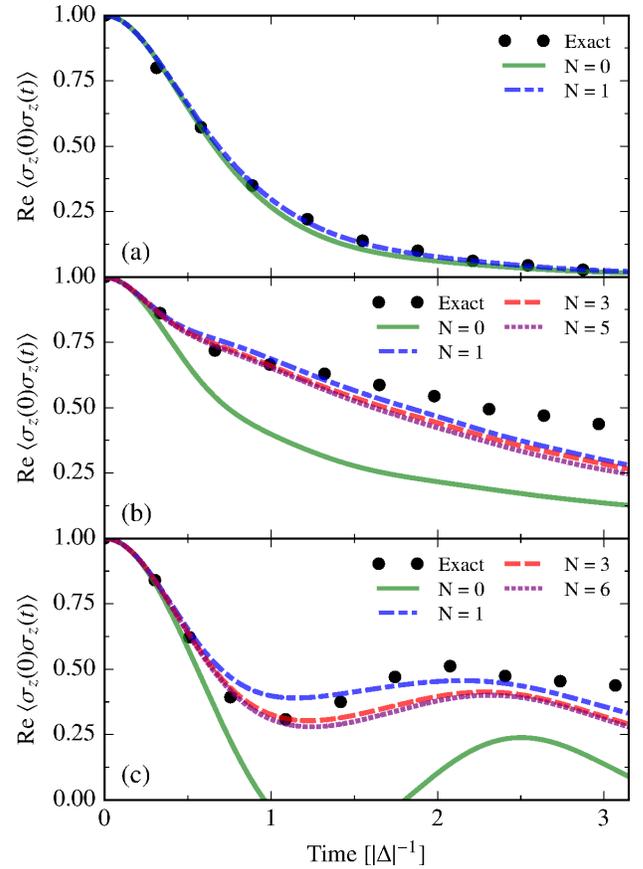} 
\caption{Representative Ehrenfest dynamics for correlation function, $\mathcal{C}_{zz}(t) = \mathrm{Re} \langle \sigma_z(0) \sigma_z(t)\rangle$, for several realizations of the unbiased ($\varepsilon = 0$) SB model. For all panels, $\Delta =-1$ and $\alpha = 1$.  For panel (a), $\omega_c = 2.5$, $\beta = 0.2$, and $\xi = 0.32$; for (b), $\omega_c = 2.5$, $\beta = 1.6$, and $\xi = 0.51$; for (c), $\omega_c = $, $\beta = 1.6$, and $\xi = 2.55$. Exact results are obtained from Ref.~\onlinecite{Mak1991}.}\label{Ch6Fig:WignerFig3}
\end{figure}

The appropriate representation of the canonical density enabled by the path integral approach presented here also facilitates the calculation of equilibrium time correlation functions.   For example, Fig.~(\ref{Ch6Fig:WignerFig3}) shows the Ehrenfest results for $\mathcal{C}_{zz}(t)$ for the unbiased SB model ($\varepsilon = 0$) obtained using representations of the canonical density that differ in the number of path integral slices employed.  Panel (a), which corresponds to a weak coupling, high temperature case, required only a minimal number of path integral slices ($N = 1$) for convergence, indicating that the system and bath are indeed approximately independent.  Also consistent with our expectations, the Ehrenfest method, which is most appropriate for systems at high temperature and weak system-bath coupling, is able to recover the exact dynamics easily.  This picture changes drastically in panels (b) and (c), which correspond to lower temperatures and greater system-bath coupling.  In these cases both the Ehrenfest method and the crude approximation for the density operator that treats the system and bath as approximately independent break down.  For these panels, the number of path integral steps necessary for the convergence of the dynamics were $N = 5$ and $6$, respectively.  It is noteworthy that the accurate rendering of the equilibrium density operator resulted in improved accuracy for the dynamics for longer times, correctly capturing the slow relaxation in panels (b) and (c), as well as the short-time behavior up to $t = \Delta^{-1}$ quantitatively. Finally, we emphasize again that our scheme for the representation of the canonical distribution can be easily incorporated into other quasi- and semi-classical schemes; we have used the Ehrenfest method to illustrate the advantages of the current approach.

\section{Conclusions}
\label{Ch6Sec:Conclusions}	

In this work we have derived an expression for the partial-Wigner transformed canonical density operator of the SB model which can be made arbitrarily accurate with increasing number of path integral slices, $N$.  This approach can be used for the evaluation of thermodynamic averages and in conjunction with quasi- and semi-classical evolution methods for the calculation of equilibrium time correlation functions. Importantly, the current work permits the systematic testing of common approximations to the quantum canonical distribution function (e.g., the thermal Gaussian and FK-LPI approaches).  Moreover, the generalization of the procedure presented here to an $M$-level system coupled linearly to a harmonic bath is straightforward.  

We have demonstrated the feasibility of the method in the calculation of thermodynamic averages for the spin and bath variables of the SB model, showing their dependence throughout parameter space.  Using the current approach with the Ehrenfest method, we have illustrated the sensitivity of the calculated dynamics to the accuracy of the representation of the canonical density operator, which is especially notable in the low temperature and high system-bath coupling regimes.  The compatibility of the expressions provided here with quasi- and semi-classical dynamical schemes opens the door to more accurate semiclassical calculations of, for instance, transport coefficients and rate constants.  We reserve the investigation of such properties for future publications.

\section*{Acknowledgements}
\label{Ch6Sec:Acknowledgements}	
D.R.R. acknowledges support from NSF Grant No. CHE-1464802. A.M.C. thanks Hsing-Ta Chen for useful conversations and Will Pfalzgraff for helpful comments on the manuscript.
	
\appendix

\section{Path Integral Treatment of the Canonical Density Operator}
\label{Ch6App:PITreatmentRho}

To derive an expression for $\rho_{a,b}(\mathbf{x} + \mathbf{s}/2, x - \mathbf{x}/2)$, necessary for the Wigner transformation of the canonical density operator in Eq.~(\ref{Ch6Eq:SchematicRho}), we first obtain expressions for the matrix elements of the Boltzmann factor using the path integral procedure outlined in Sec.~\ref{Ch6Subsec:CanonicalDensity}. Specifically, we use the Trotter decomposition in Eq.~(\ref{Ch6Eq:Trotter}) and introduce resolutions of the identity in the system and bath subspaces, $\mathbf{1}_S = \sum_{a} \ket{a}\bra{a}$ and $\mathbf{1}_B = \int d\mathbf{q} \ket{\mathbf{q}}\bra{\mathbf{q}}$, so that the matrix elements of the Boltzmann factor can be rewritten as
	\begin{equation}\label{Ch6Eq:FPI}
	\begin{split}
	F_{k_N, k_0}(\mathbf{Q}_N, \mathbf{Q}_0) &= \bra{\mathbf{Q}_N} \bra{k_N} e^{-\beta H} \ket{k_0}\ket{\mathbf{Q}_0}\\
	&\approx \sum_{\{ k_1, ..., k_{N-1}\}} \tilde{\mathcal{S}}_{k_N, k_0} \tilde{\mathcal{B}}_{k_N, k_0}(\mathbf{Q}_N, \mathbf{Q}_0),
	\end{split}
	\end{equation}
	where
	\begin{align}
	\tilde{\mathcal{S}}_{k_N, k_0} &= \prod_{j = 1}^{N} \bra{k_j}e^{-\beta H_{ad}/N} \ket{k_{j-1}}, \label{Ch6Eq:SPI}\\
	\tilde{\mathcal{B}}_{k_N, k_0}&(\mathbf{Q}_N, \mathbf{Q}_0) = \int d\mathbf{Q}_{1}... d\mathbf{Q}_{N-1} \nonumber \\
	&\qquad \prod_{j = 1}^{N} \bra{\mathbf{Q}_j}e^{-\beta H^{k_{j}}_{na}/2N}e^{-\beta H^{k_{j-1}}_{na}/2N} \ket{\mathbf{Q}_{j-1}}, \label{Ch6Eq:BathPartPI}
	\end{align}
	\begin{align}
	H^{k_j}_{na} &= \frac{1}{2} \sum_l \Big[\hat{P}_l^2 + \omega_l^2 (\hat{Q}_l - b_{k_j}^{(l)})^2\Big], \\
	b_{k_j}^{(l)} &= (-1)^{k_j}\alpha c_l/\omega_l^2.
	\end{align}
In this notation, 
	\begin{equation}
	\rho_{a,b}(\mathbf{x} + \mathbf{s}/2, x - \mathbf{x}/2) = \frac{F_{a,b}(\mathbf{x} + \mathbf{s}/2, \mathbf{x} - \mathbf{s}/2)}{Z},
	\end{equation}	
	\begin{equation}\label{Ch6Eq:InstZ}
	Z =  \sum_{a}\int d\mathbf{x}\ F_{a,a}(\mathbf{x}, \mathbf{x}).
	\end{equation}	
	
	The the path integral unit, $\bra{\mathbf{Q_{n}}}e^{-\beta H_{nd}^{k_n} /2N }  e^{-\gamma H^{k_{m}}_{nd} /2N }\ket{\mathbf{Q}_{m}}$, in Eq.~(\ref{Ch6Eq:BathPartPI}) takes the following form,\cite{Makri1992, Makarov1994}
\begin{widetext}
	\begin{equation}\label{Ch6Eq:PIunit}
	\begin{split}
	\bra{\mathbf{Q_{n}}}e^{-\beta H_{nd}^{k_n} /2N }  &e^{-\gamma H^{k_{m}}_{nd} /2N }\ket{\mathbf{Q}_{m}} = \prod_{l=1}^{f} \sqrt{\frac{\omega_l }{2\pi\sinh(2\theta_l)}} \exp\Bigg[ -\frac{\omega_l}{2\sinh(2\theta_l)}\Big[ [(\delta Q^{(l)}_n)^2 + (\delta Q^{(l)}_m)^2] \cosh(2\theta_l) \\
	& \hspace{10em} + 2\cosh(\theta_l)(\delta Q^{(l)}_n  -  \delta Q^{(l)}_m)\Delta b^{(l)}_{nm}   - 2 \delta Q^{(l)}_n \delta Q^{(l)}_m + (\Delta b^{(l)}_{nm})^2 \cosh^2(\theta_l) \Big] \Bigg],
	\end{split}	
	\end{equation}
where $\delta Q^{(l)}_{n} = Q^{(l)} - b_{n}^{(l)}$ is the difference between the coordinate of the $l^{th}$ harmonic oscillator and its displacement due to the system-bath coupling, $\Delta b^{(l)}_{nm} = b^{(l)}_{n} - b^{(l)}_{m}$, and $\theta_l = \beta \omega_l/2N$.  

With the previous definitions, it is possible to obtain the following expression
	\begin{equation}\label{Ch6Eq:ExplicitBPI}
	\begin{split}
	\tilde{\mathcal{B}}_{a, b}&(\mathbf{x} + \mathbf{s}/2, x - \mathbf{x}/2) = \prod_{l = 1}^{f}\sqrt{\frac{\omega_l}{\pi \mathrm{det}[\mathbf{A}^{(l)}]}}\exp\Bigg[ -\gamma_x^{(l)} (x_l + \tilde{\kappa}_x^{(l)})^2 -\gamma_p^{(l)} s_l^2 + \tilde{\kappa}_p^{(l)}s_l - \tilde{\Lambda}^{(l)}\Bigg], 
	\end{split}
	\end{equation}
where $\mathbf{A}^{(l)}$ is a tridiagonal $N-1 \times N-1$ matrix whose diagonal and off-diagonal entries are equal to $2$ and $-\mathrm{sech}(2\theta_l)$, respectively. For $N < 2$, $\mathrm{det}[\mathbf{A}^{(l)}] = 1$. The path-dependent quantities above (marked by a tilde) take the forms, 
	\begin{equation}
	\begin{split}
	\tilde{\kappa}_{p}^{(l)} =  \left\{
      \begin{array}{lr}
       -\frac{\omega_l}{2\tanh(2\theta_l)}\Bigg[\frac{\cosh(\theta_l)}{\cosh(2\theta_l)}[(\Delta b^{(l)}_{N,N-1} +\Delta b^{(l)}_{1,0})-(\tilde{\boldsymbol{\delta}}^{(l)}_{N-1}-\tilde{\boldsymbol{\delta}}^{(l)}_1)] - \eta_l \Delta b^{(l)}_{N,0} \Bigg] \quad &: \quad N \geq 2,\\
       -\frac{\omega_l}{\tanh(2\theta_l)}\Bigg[\frac{\cosh(\theta_l)}{\cosh(2\theta_l)}\Bigg]\Delta b^{(l)}_{1,0} [1 - \cosh(\theta_l)]  &: \quad N=1,
      \end{array}
    \right.
	\end{split}
	\end{equation}
	\begin{equation}
	\begin{split}
	\tilde{\kappa}_{x}^{(l)} =  \left\{
      \begin{array}{lr}
       \frac{\cosh(\theta_l)}{2\cosh(2\theta_l)}\frac{  [(\Delta b^{(l)}_{N,N-1} - \Delta b^{(l)}_{1,0}) - (\tilde{\boldsymbol{\delta}}^{(l)}_{N-1}+\tilde{\boldsymbol{\delta}}^{(l)}_1)]}{\nu_l} - \frac{b^{(l)}_N + b^{(l)}_0}{2} \quad &: \quad N \geq 2,\\
       - \frac{b^{(l)}_N + b^{(l)}_0}{2}  &: \quad N=1,
      \end{array}
    \right. 
	\end{split}
	\end{equation}
	\begin{equation}
	\begin{split}
	\tilde{\Lambda}^{(l)} =  \left\{
      \begin{array}{lr}
       \frac{\omega_l}{4 \tanh(2\theta_l)} \Bigg[  \frac{1 + \cosh(2\theta_l)}{\cosh(2\theta_l)}\Big[\sum_{j = 1}^{N} [\tilde{\boldsymbol{\delta}} b^{(l)}_{j,j-1}]^2 -\frac{\tilde{\mathbf{j}}_l^T \cdot \mathbf{A}_l^{-1} \cdot \tilde{\mathbf{j}}_l}{\cosh(2\theta_l)}\Big]  - 2 \frac{\cosh(\theta_l)}{\cosh(2\theta_l)}\Big[(\Delta b^{(l)}_{N,N-1}+\Delta b^{(l)}_{1,0})-(\tilde{\boldsymbol{\delta}}^{(l)}_{N-1}-\tilde{\boldsymbol{\delta}}^{(l)}_1)\Big]\Delta b^{(l)}_{N,0} \quad &  \\
       \qquad \qquad  \qquad - \Big[\frac{\cosh(\theta_l)}{\cosh(2\theta_l)}\Big]^2\frac{ [(\Delta b^{(l)}_{N,N-1} - \Delta b^{(l)}_{1,0}) - (\tilde{\boldsymbol{\delta}}^{(l)}_{N-1}+\tilde{\boldsymbol{\delta}}^{(l)}_1)]^2}{\nu_l} + \eta_l [\Delta b^{(l)}_{N,0}]^2\Bigg] \quad &: \quad N \geq 2,\\
       \frac{\omega_l}{\tanh(2\theta_l)}\Bigg[\frac{\cosh(\theta_l)}{\cosh(2\theta_l)}\Bigg][\Delta b^{(l)}_{1,0}]^2 [1 - \cosh(\theta_l)]  &: \quad N=1,
       \end{array}
    \right. 
	\end{split}
	\end{equation}
\end{widetext}
	where
	\begin{equation}
	\begin{split}
	\tilde{\mathbf{j}}_l &= \left[ \begin{array}{c}
\Delta b^{(l)}_{21} - \Delta b^{(l)}_{10}   \\
\Delta b^{(l)}_{32} - \Delta b^{(l)}_{21}  \\
\vdots \\
\Delta b^{(l)}_{N,N-1} - \Delta b^{(l)}_{N-1,N-2}   \end{array}\right],   
\end{split}
	\end{equation}
and $\tilde{\boldsymbol{\delta}}^{(l)} = \mathbf{j}_{l}^T \cdot \mathbf{A}_l^{-1} /\cosh(\theta_l)$. Also, when $N = 0$, $\tilde{\kappa}_p^{(l)} = \tilde{\kappa}_x^{(l)} = \tilde{\Lambda}^{(l)} = 0$.

The path-independent quantities take the following forms,
	\begin{align}
	\eta_l &= 1-\frac{[\mathbf{A}_{l}^{-1}]_{1,1} - [\mathbf{A}^{-1}_l]_{1,N-1}}{\cosh^2(2\theta_l)},\\
	\nu_l &= 1-\frac{[\mathbf{A}_{l}^{-1}]_{1,1} + [\mathbf{A}_l^{-1}]_{1,N-1}}{\cosh^2(2\theta_l)},\\
	\gamma^{(l)}_{p}  &= \frac{\tanh(2\theta_l)}{\omega_l \eta_l}, \\
	\gamma^{(l)}_{x} &= \frac{\omega_l \nu_l }{\tanh(2\theta_l)}. 
	\end{align}
For $N < 2$, $\eta_l = \nu_l = 1$. 
		
Substituting Eqs.~(\ref{Ch6Eq:ExplicitBPI}) and (\ref{Ch6Eq:SPI}) into Eq.~(\ref{Ch6Eq:FPI}), setting $\mathbf{s} = 0$, and performing the integration in Eq.~(\ref{Ch6Eq:InstZ}) leads to the following expression for the partition function, 
\begin{equation}\label{Ch6Eq:PartitionFunction}
	\begin{split}
	Z &=\Big[\sum_{a} \mathcal{W}_{a,a}\Big] \prod_{l =1}^{f} \Big[ 2\cosh(2\theta_l) \eta_l \mathrm{det}[\mathbf{A}_{l}] \Big]^{-1/2} , 
	\end{split}
	\end{equation}	
where the path-dependent weights take the form $\tilde{W}_{a,b} = \tilde{\mathcal{S}}_{a,b}  \exp[-\sum_l \tilde{\Lambda}_{a,b}^{(l)}]$, and $\mathcal{W}_{a,b} = \sum_{paths} \tilde{W}_{a,b}$.

One final integration over $\mathbf{s}$ in Eq.~(\ref{Ch6Eq:SchematicRho}) leads to the following expressions for the partial bath distribution and normalization factor, 
	\begin{equation}
	\begin{split}
	\mathcal{R}_{a,b}^W(\mathbf{x}, \mathbf{p}) &= \mathcal{N} \sum_{\{ k_1, ..., k_{N-1} \}} \frac{\tilde{W}_{a,b}}{\mathcal{W}_{a,b}}  \\
	&  \prod_{l = 1}^{f} \exp\Big[-\gamma_p^{(l)}( p_l + i\tilde{\kappa}_p^{(l)})^2 -\gamma_x^{(l)} (x_l + \tilde{\kappa}_x^{(l)})^2 \Big],
	\end{split}
	\end{equation}
	\begin{equation}
	\begin{split}
	N_{ab} &=\frac{\mathcal{W}_{a,b}}{\sum_a  \mathcal{W}_{a,a}},
	\end{split}
	\end{equation}
	 \begin{equation}
	 \mathcal{N} = \Bigg[ \prod_{l = 1}^{f} \frac{\sqrt{\nu_l/\eta_l}}{\pi }\Bigg].
	 \end{equation}
	 
Clearly, the equations derived above have explicitly used the fact that the bath can consists of independent oscillator.  To ensure compatibility with the second line of Eq.~(\ref{Ch6Eq:OhmicSD}), we use the approach outlined in Ref.~\onlinecite{Craig2005} which allows us to decompose the spectral density into $f$ oscillators. The frequency of the $k^{th}$ oscillator takes the form, 
	\begin{equation}
	\omega_k = -\omega_c \ln\Bigg[ \frac{k - \frac{1}{2}}{f} \Bigg], 
	\end{equation}
and the coupling constant, 
	\begin{equation}
	c_k = \omega_k  \Bigg[\frac{\xi \omega_c}{f}\Bigg]^{1/2} .
	\end{equation}
	For the results shown here, we used $f = 200 - 300$ oscillators.
	
\section{Ehrenfest method}
\label{Ch6App:Ehrenfest}

The Ehrenfest method \cite{Gerber1982, Stock1995, Tully1998a, Grunwald2009} is a wavefunction-based approach where the system (bath) evolves in the mean field of the bath (system).  In addition, this scheme assumes that the bath dynamics are correctly captured by classical mechanics. One may rigorously formulate the Ehrenfest method by first performing a partial Wigner transform with respect to the bath degrees of freedom of the dynamical object to be calculated, e.g., nonequilibrium average or time correlation function,
	\begin{equation}\label{Ch6Eq:EhGenC}
	\begin{split}
	C_{AB}(t) &= \mathrm{Tr}[\mathcal{A}_S(0) \mathcal{A}_B(0) \mathcal{B}_S(t) \mathcal{B}_B(t)]\\
	&\approx \int d\mathbf{x}\mathbf{p}\ \mathcal{A}_B ^W \mathcal{B}^W_B(t)\mathrm{Tr}_S[ \mathcal{A}_S(0)   \mathcal{B}_S(t)]
	\end{split}   
	\end{equation}
where $X_S$ ($X_B$) is a generic system (bath) operator.  

The heart of the approximation in the Ehrenfest method lies in the dynamical treatment of the operators.  In this scheme, the time-dependence is given by the equations of motion for the system and bath.  In the case of the system, the wavefunction is evolved via the quantum Liouville equation under the influence of a modified Hamiltonian, 
	\begin{equation}
	\frac{d}{dt} \rho_S(t) = -i[H_{S}^{Eh}, \rho(t)], 
	\end{equation}
where 
	\begin{equation}
	H_{S}^{Eh}(t) = [\varepsilon + \lambda^{cl}(t)] \sigma_z + \Delta \sigma_x,
	\end{equation}
is the modified system Hamiltonian and $\lambda^{cl}(t) = \alpha \sum_k c_k Q_k(t)$ is the classical fluctuation in the bias provided by the classical treatment of the bath.  Here, $\rho_S(0)$ is the initial density matrix for the system.  In Eq.~(\ref{Ch6Eq:EhGenC}), this corresponds to operator $\mathcal{A}(0)$.  Since the Ehrenfest is a wavefunction based method, initial conditions corresponding to coherences, $\rho_S(0) = \ket{i}\bra{j}$ where $i \neq j$, must first be sampled correctly for the Ehrenfest method to yield appropriate results. Details regarding the generation may be found in Ref.~\onlinecite{Montoya2016a}. 

The equations of motion for the bath variables are given by the classical Hamilton's equations subject to the time-dependent Hamiltonian, 
	\begin{align}
	\frac{dP_k}{dt}  &= - \frac{\partial H_{B}^{Eh}}{\partial Q_k},\\
	\frac{dQ_k}{dt}  &=  \frac{\partial H_{B}^{Eh}}{\partial P_k},
	\end{align}	 
where 
\begin{equation}
	H_{B}^{Eh}(t) = \frac{1}{2} \sum_k \Big[ P_k^2 + \omega_k^2 Q_k + 2\alpha \bar{\sigma}_z(t) c_k Q_k \Big], 
	\end{equation}
and $\bar{\sigma}_z(t) = \mathrm{Tr}_S[\rho_S(t) \sigma_z]$

Given the previous considerations, $\mathcal{C}_{zz}(t)$ takes the form,
	\begin{equation}\label{Ch6Eq:EhCzz}
	\begin{split}
	C_{zz}(t) &= \mathrm{Re}\ \mathrm{Tr}[\rho \sigma_z(0)\sigma_z(t)]\\
	&= \mathrm{Re} \sum_{a} \Bigg[ N_{a,1}\int d\mathbf{x} d\mathbf{p}\ \mathcal{R}_{a,1}^W(\mathbf{x}, \mathbf{p})\mathrm{Tr}_S[\ket{a}\bra{1} \sigma_z(t)]\\
	&\qquad   + N_{a,2}\int d\mathbf{x} \mathbf{p}\ \mathcal{R}_{a,2}^W(\mathbf{x}, \mathbf{p})\mathrm{Tr}_S[\ket{a}\bra{1} \sigma_z(t)]\Bigg]. 
	\end{split}  
	\end{equation}
To calculate $\mathcal{C}_{zz}(t)$, a second-order Runge-Kutta scheme was implemented.  During individual time steps, $\bar{\sigma}_z(t)$ is kept constant for the evolution of the bath, while $\lambda^{cl}(t)$ is kept constant during the evolution of the system. Over a half time step, the equations for the classical variables take the forms,
	\begin{equation}
    \begin{split}
    Q_{k}\left(t + \frac{\delta t}{2} \right) &= \gamma_{k}(t)\cos\left( \frac{\omega_{k} \delta t}{2} \right) - \frac{\alpha c_{k}}{\omega_{k}^2}\bar{\sigma_z}(t)\\
    &\qquad \qquad + \frac{P_{k}(t)}{\omega_{k}}\cos\left(\frac{\omega_{k} \delta t}{2}\right) ,
    \end{split}
    \end{equation}
and 

    \begin{equation}
    \begin{split}
    P_{k}\left(t + \frac{\delta t}{2} \right) &= P_{k}(t)\cos\left(\frac{\omega_{k} \delta t}{2} \right) + \omega_{k}\gamma_{k}(t)\sin\left(\frac{\omega_{k} \delta t}{2}\right),
    \end{split}
    \end{equation}
where
    \begin{equation}
    \gamma_{k}(t) =  Q_{k}(t) + \frac{\alpha c_{k}}{\omega_{k}^2}\bar{\sigma_z}(t).
    \end{equation} 
Convergence for the correlation functions was achieved using $\sim 5 \times 10^4 - 10^5$ trajectories.

%


\end{document}